\documentclass[useAMS,usenatbib]{mn2e}

\usepackage{graphicx}
\usepackage{multirow}
\usepackage{amssymb}
\usepackage{aas_macros}
\usepackage{times}

\def \spitzer {{\it Spitzer}}
\def \wise {{\it WISE}}
\def \oiii {[O{\sc iii}]}
\def \nii {[N{\sc ii}]}
\def \sii {[S{\sc ii}]}
\def \ha {H$\alpha$}
\def \hb {H$\beta$}

\title[The Nature of Obscuration in AGN]{The Nature of Obscuration in AGN: I. Insights from Host Galaxies}

\author[L.~Shao et al.]{Li Shao,$^1$\thanks{E-mail: lishao@mpa-garching.mpg.de} Guinevere Kauffmann,$^1$ Cheng Li$^2$, Jing Wang$^1$, Timothy M. Heckman$^3$\\
$^1$ Max Planck Institute for Astrophysics, Karl-Schwarzschild-Str. 1, Garching, 85748, Germany\\
$^2$ Partner Group of the Max Planck Institute for Astrophysics at the Shanghai Astronomical Observatory and Key Laboratory \\
for Research in Galaxies and Cosmology of Chinese Academy of Sciences, Nandan Road 80, Shanghai 200030, China\\
$^3$ Center for Astrophysical Sciences, Department of Physics and Astronomy, Johns Hopkins University, Baltimore, MD 21218, USA\\}

\begin{document}

\date{Accepted on Sep 25}

\pagerange{\pageref{firstpage}--\pageref{lastpage}} \pubyear{2013}

\maketitle

\label{firstpage}

\begin{abstract}
  We analyze a sample of 30,000 nearby obscured AGNs with optical spectra from SDSS and mid-IR photometry from WISE. Our aim is to investigate the AGN host galaxy properties with mid-IR luminosities as AGN activity indicator, and to compare with previous studies based on \oiii\ emission lines. First we find that the [3.4] - [4.6] colour has weak dependence on host stellar age, but strong dependence on AGN activity. We then use a ``pair-matching'' technique to subtract the host 4.6 micron contribution. By combining Seyferts with a sample of SDSS quasars at $z<0.7$, we show that the \oiii\ and the intrinsic AGN 4.6 micron luminosities correlate roughly linearly over 4 orders of magnitude, but with substantial scatter. We also compare the {\em partition functions} of the total integrated 4.6 micron and \oiii\ line luminosities from Seyferts and a sub-population of LINERs with significant nuclear 4.6 micron emission, as function of a variety of host galaxy properties, finding that they are identical. We conclude, therefore, that \oiii\ as an AGN indicator shows no particular biases as compared to the 4.6 micron luminosity. Our results also demonstrate that some LINERs do fit in with the expectations of the simple Unified Model.
\end{abstract}


\section{Introduction}
\label{sec:intro}

The energy output of active galactic nuclei (AGNs) is thought to be a good probe of black hole (BH) growth history in the Universe. In principle, observing the radiation from the innermost part of the accretion system yields a direct estimate of the total mass of gas swallowed by the central black hole. Luminous type 1 quasars clearly exhibit a featureless continuum which originates from the hot accretion disk. However, in many type 2 AGNs it is not possible to observe the accretion disk because the radiation is absorbed by the surrounding gas. These AGNs lack the power-law continuum and broad optical emission lines from the accretion disk. Instead, the spectral energy distribution (SED) is dominated by stellar emission from the host galaxy. The presence of an actively accreting black hole is manifested by high ionization {\em narrow} emission lines arising from gas a few hundred parsecs away from the black hole that is being irradiated by the accretion disk. The unification model of AGNs \citep{1993ARA&A..31..473A,1995PASP..107..803U} postulates that an anistrotropically distributed collection of absorbing clouds (often referred to as the
``torus'') can lead to a natural explanation for varying types of AGNs with very different SEDs.

The fact that the accretion disk is obscured in type 2 AGNs means that we can study the host galaxy due to greatly reduced contrast between core radiation and outer stellar emission. There have been longstanding efforts to answer key questions, such as what kind of galaxies host AGN and what triggers the accretion onto the black hole \citep[e.g.,][]{1980A&A....87..152H,1997ApJS..112..315H,1999MNRAS.308..377M}. By investigating a large sample of type 2 AGNs, \citet{2003MNRAS.346.1055K} found that local AGNs are mostly hosted by galaxies with stellar masses greater than $10^{10}\,\mathrm{M_\odot}$ and with stellar surface densities greater than $3\times10^8\,\mathrm{M_\odot}$. Black hole growth is currently occurring in low mass black holes located in low mass bulges, which are also still forming stars at present \citep{2004ApJ...613..109H}. The ratio between the current black hole growth rate by accretion and bulge growth rate by star formation is $\sim$0.001, consistent with the observed $M_\mathrm{BH}-M_\mathrm{bulge}$ relation \citep[e.g.,][]{2003ApJ...589L..21M,2004ApJ...604L..89H}.

Further inspection of the relation between star formation in the host galaxy and the accretion rate onto the black hole suggests that there are two modes of AGN accretion \citep{2009MNRAS.397..135K}. The majority of AGN hosts have little star formation and old stellar populations. These AGNs may be fed by stellar winds from evolved stars. Their inferred accretion rates are low on average, but the duty cycle for this kind of activity is high \citep{2003MNRAS.346.1055K,2009MNRAS.397..135K}. It has been suggested that a substantial fraction of such objects may not be ``true'' AGN, because the low ionization lines could be produced by radiation from evolved stars \citep{2011MNRAS.413.1687C,2012ApJ...747...61Y}.

A small fraction of AGN hosts show clear recent central star formation \citep{2003MNRAS.346.1055K,2007MNRAS.381..543W,2007ApJS..173..357K,2010MNRAS.406L..40L}. These AGNs have higher accretion rates on average and occur in galaxies where black holes and bulges are growing simultaneously. It has been suggested that a plentiful cold gas supply is the common source for optical AGN activity and central star formation \citep{2007ApJS..173..357K,2009MNRAS.397..135K}. 

Obscuration makes the estimation of accretion rate difficult. Because the accretion disk continuum is hidden, it is necessary to find indirect indicators to represent the accretion power. Usually these indicators measure the amount of ''reprocessed'' radiation. In the optical, emission lines from the narrow line region are commonly used, especially the \oiii5007 line \citep[e.g.][]{2004ApJ...613..109H}. The narrow line emission arises from gas that extends over much larger scales than the accretion disk; the emission region is observed to extend over scales of a few hundred parsecs in type 2 AGNs. The hard X-ray emission from the corona surrounding the accretion disk is another popular indicator of black hole accretion. At longer wavelengths, the optical/ultraviolet radiation absorbed by the torus is re-emitted in mid-infrared, and the mid-IR luminosity is expected to be correlated with accretion rate. Some mid-IR high-ionization emission lines are also used as AGN activity indicators \citep{2008ApJ...674L...9D,2008ApJ...682...94M}. At even longer wavelengths, radio emission from the jet is another possible indicator of black hole accretion rate.

Generally speaking, except for lower luminosity FR-I type radio AGNs, these indicators are correlated with each other \citep[e.g.,][]{2005ApJ...634..161H,2008ApJ...682...94M,2010ApJ...720..786L,2010A&A...515A..23H,2012ApJ...754...45I}. However, clear discrepancies have also been reported, leading to the worry that any AGN survey carried out at only one wavelength may bias our understanding. Lower-luminosity FR-I type radio AGNs are clearly distinct from optical AGNs because they occur in more massive galaxies \citep{2005MNRAS.362...25B}.

The AGN samples from deep X-ray surveys are considered to be close to complete \citep[see][for a review]{2012NewAR..56...93A}, yet \citet{2005ApJ...634..161H} raise the possibility that many \oiii\ bright AGNs are missed from hard X-ray surveys. The analysis of the cosmic X-ray background suggests the existence of a class of AGNs with very high column density along the line of sight, or so called Compton-thick AGNs \citep{2007A&A...463...79G,2009ApJ...696..110T}. It is commonly agreed that this kind of objects are missed even by the deepest X-ray survey but (at least) some of them are identified in optical/IR observations \citep{2005ApJ...634..161H,2006A&A...455..173P,2008ApJ...682...94M,2011MNRAS.411.1231G}. Similarly, optical identification techniques may miss some strongly accreting black holes. \citet{2006A&A...453..525N} and \citet{2010ApJ...722..212T} find that the \oiii/$L_X$ ratio decreases with increasing X-ray luminosity, which implies that some X-ray AGNs are probably not identified in optical surveys. \citet{2010MNRAS.406..597G} also argues that optical surveys are missing the accretion around low mass black holes.

Both X-ray and optical techniques are clearly affected by extinction/absorption. However, short-wavelength radiation that is absorbed by dust, will be re-radiated at IR wavelengths, so ``missing'' objects should be recovered in IR-selected surveys. Previous work has claimed that the number of detected AGNs is greatly increased in IR surveys \citep{2008ApJ...672...94F,2008ApJ...687..111D}. Nevertheless it is also reported that IR-selected AGNs are biased against weak and type 2 AGNs \citep{2008ApJ...680..130C}. Some X-ray selected AGNs have been found to have IR colours consistent with pure star-forming galaxies, suggesting that IR colour-selection techniques will miss accreting black holes in host galaxies with strong star formation \citep{2009A&A...507.1277B}.

In order to understand fully how black holes grow, we need to study AGN at multiple wavelengths. Some efforts have been made to unify the view of AGNs from different wavelengths based on the data in some specific sky region with intensive multi-wavelength coverage. \citet{2009ApJ...696..891H} compare X-ray, IR and radio-selected AGNs using data from the AGN and Galaxy Evolution Survey (AGES). They find that the X-ray and IR-selected AGNs are similar in terms of host galaxy properties and clustering strength, while radio AGNs are clearly more strongly clustered. Optically-faint AGNs have harder X-ray spectra than typical unabsorbed AGNs. Differences between X-ray and IR AGN hosts are mainly attributed to the slight stellar/BH mass difference. \citet{2012MNRAS.427.3103B} compile a catalogue of obscured and unobscured AGNs from the Cosmic Evolution Survey (COSMOS) sky region. They perform detailed multi-wavelength SED decomposition into stellar and AGN components, estimating the AGN accretion rate and host properties simultaneously. They confirm that the AGN fraction is higher in more massive galaxies, independent of accretion rate.

These studies, however, are based on relatively small samples. In this paper, we will perform an analysis of a large sample of local obscured AGNs, based on the data from the Sloan Digital Sky Survey \citep[SDSS,][]{2000AJ....120.1579Y} and the Wide-field Infrared Survey Explorer \citep[\wise,][]{2010AJ....140.1868W}. In section \ref{sec:data} we will describe the organization of the data from the SDSS/\wise\ database and how we construct our local galaxy sample. In section \ref{sec:sdss} we will analyze the \wise\ colour properties of local galaxies, including passive galaxies, star forming galaxies and AGNs. We explore how \wise\ colours change as a function of position in the 2-dimensional plane of 4000 \AA\break strength and black hole accretion rate, as quantified by the extinction-corrected \oiii\ line luminosity scaled by the black hole mass of the galaxy. This allows us to specify which colours are sensitive primarily to star formation, and which to AGN activity.

We then compare the host galaxies of IR-selected AGN with those of optically-selected AGN in section \ref{sec:optagn}. We do this by examining how the total IR emission from the torus in nearby AGN is distributed across galaxies as a function of global properties such as stellar mass, stellar mass surface density, concentration, and specific star formation rate (sSFR). We compare these IR emissivity "distribution functions" to results obtained from integrating up the total \oiii\ emission from narrow-line regions \citep{2004ApJ...613..109H}. In section \ref{sec:iragn}, we examine AGNs that are clearly identified at IR wavelengths, but not at optical wavelengths, and ask whether and how they are different from the rest of the population. Finally we will discuss our results and give a brief summary in section \ref{sec:discussion}.

\section{Data and Sample}
\label{sec:data}

\subsection{Matching the SDSS spectroscopic and \wise\ photometric catalogues}
\label{ssec:data_catalogue}

Our parent sample includes all galaxies from the MPA/JHU SDSS DR7 catalogue\footnote{http://www.mpa-garching.mpg.de/SDSS/} with magnitudes in the range $14.5<r<17.6$. The $r>14.5$ cut is used to remove the nearby galaxies with large angular sizes, because both the SDSS and \wise\ pipeline photometry will fail for highly extended sources. We limit our analysis to the redshift range $0.02<z<0.21$, so we do not have to worry about evolutionary effects within our sample. The typical $r$-band 50\% light radii $R_{50}$ of our galaxies in this redshift range are $\sim2.5$ arcsecs, and only few (86, $\sim0.04\%$) objects have $R_{50}$ larger than 10 arcsecs. The stellar masses, which are directly obtained from the MPA/JHU catalogue, are estimated by fitting broadband SEDs to a library of synthesis models \citep[e.g.][]{2007ApJS..173..267S}. The uncertainty of the stellar mass estimation is $\sim0.1$ dex. Given the fact that the majority of AGNs are hosted by massive galaxies \citep{2003MNRAS.346.1055K}, we limit our study to galaxies with $\log(M_*/\mathrm{M_\odot})>9.8$.

\begin{figure}
 \centering
 \includegraphics[width=8cm]{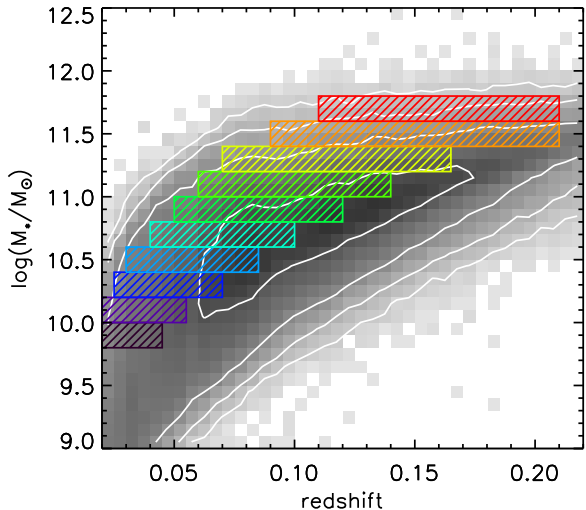}
 \caption{The redshift versus stellar mass. The background grey-coded histogram and the white contours show the distribution of the whole SDSS spectroscopic sample with a $14.5<r<17.6$ cut on the $r$-band model magnitude. The coloured boxes indicate the different stellar mass bins and redshift ranges for the volume-limited subsamples we use.\label{fig:zmass}}
\end{figure}

\begin{table}
 \centering
 \begin{tabular}{c|cr}
  $\log(M_*/\mathrm{M_\odot})$ & \multicolumn{1}{c}{Redshift} & \multicolumn{1}{c}{Number} \\ \hline
  9.8 - 10.0 & 0.020 - 0.045 & 5673 \\
  10.0 - 10.2 & 0.020 - 0.055 & 9661 \\
  10.2 - 10.4 & 0.025 - 0.070 & 18066 \\
  10.4 - 10.6 & 0.030 - 0.085 & 30201 \\
  10.6 - 10.8 & 0.040 - 0.100 & 36404 \\
  10.8 - 11.0 & 0.050 - 0.120 & 41500 \\
  11.0 - 11.2 & 0.060 - 0.140 & 36133 \\
  11.2 - 11.4 & 0.070 - 0.165 & 23083 \\
  11.4 - 11.6 & 0.090 - 0.210 & 13212 \\
  11.6 - 11.8 & 0.110 - 0.210 & 2338 \\ \hline
 \end{tabular}
 \caption{The stellar mass bins for the sub-samples that make up volume-limited sample. The corresponding redshift ranges and the number of sources in each bin are also listed.\label{tab:sample_cuts}}
\end{table}

We first define a sample of galaxies that is ``complete'' in stellar mass. We do this by dividing galaxies with stellar masses $\log(M_*/\mathrm{M_\odot})>9.8$ into bins of width 0.2 dex in $\log(M_*)$ and evaluating the redshift interval over which all such galaxies are detected in the SDSS spectroscopic sample. This is illustrated in detail in Figure \ref{fig:zmass}. The upper redshift limits show where the $r$-band flux limit means we can no longer detect all galaxies in the given stellar mass range. Our cuts are similar to the sample definition adopted by \citet{2010MNRAS.404.1231V} in their Figure 5. The lower redshift limits are imposed to avoid galaxies with angular sizes that are too large for the SDSS and \wise\ catalogue photometry to be reliable. There are a total of 216272 SDSS sources in these volume-limited samples. We list the sample details in Table \ref{tab:sample_cuts}.

The SDSS galaxies are matched to the \wise\ catalogue within a search radius of 3" from the optical position. Given the small astrometry errors for both the SDSS and the \wise\ catalogues, the probability of mismatches is negligible. In order to get reliable flux measurements, we only use the fluxes with signal-to-noise ratio larger than 3. However, a large number of sources are extended sources. In this case, the default \wise\ pipeline profile-fitting photometry will underestimate the the total flux, so we use the total magnitudes derived from elliptical aperture photometry instead. The parameters of the elliptical apertures, such as axis ratios and position angles, are not derived from the \wise\ images themselves, but are taken from the 2MASS Extended Source Catalog \citep{2006AJ....131.1163S}. For more detailed discussion of the elliptical aperture photometry, please see section IV.4.c and VI.3.e of \citet{2012wise.rept....1C}.

The \wise\ Vega magnitudes are converted to AB magnitudes and absolute fluxes. The SED shape affects the conversion from the total flux to monochromatic flux, so additional colour corrections are necessary. \citet{2010AJ....140.1868W} tabulate the relation between spectral shape and colour correction factors. Here for simplicity, we only use the table entries for power law forms $F \propto \nu^\alpha$, and we interpolate between the \wise\ colours provided in the table to derive our corrections, which are typically less than 3\% for the 3.6, 4.6 and 22 micron bands (the 12 micron band is very broad, so the corrections can be as high as 10\%).

In the whole sample, 213789 (98.9\%) sources are detected by \wise\ in the 3.4 and 4.6 micron bands at the 3-$\sigma$ level and above. Only 54324 (25.1\%) sources have $>3\sigma$ detections in all \wise\ bands. The 22 micron band has the lowest detection rate. We thus contruct a subsample by adopting a 22 micron flux cut of 7 mJy. This flux level is where the \wise\ images with average number of scan frames will recover $\sim$95\% of ``real'' sources \citep[see section VI.5c of][for more details]{2012wise.rept....1C}. There are 21942 sources in this subsample (hereafter we call the whole volume-limited sample as S1 and this subsample as S2), and 21254 (96.9\%) of them are detected in all \wise\ bands. We use S2 to study the colour distributions of galaxies in section \ref{sec:sdss}. In Table \ref{tab:detrat} we summarize the \wise\ detection rates of our various samples.

\begin{table}
 \centering
 \begin{tabular}{l|rrr}
  Detection & \multicolumn{1}{c}{S1} & \multicolumn{1}{c}{S2} \\ \hline
  Any band & 213789 (98.85\%) & 21942 (100.00\%) \\
  W1+W2 & 209142 (96.70\%) & 21330 (97.21\%) \\
  W4 & 56697 (26.22\%) & 21942 (100.00\%) \\
  All bands & 54324 (25.12\%) & 21254 (96.86\%) \\ \hline
 \end{tabular}
 \caption{The detection rates in \wise\ bands. S1 and S2 are samples we use in this paper (see text for detail). ``W1+W2'' means detection in both 3.4 and 4.6 micron bands. ``W4'' means detection in 22 micron band.\label{tab:detrat}}
\end{table}

\begin{figure}
 \centering
 \includegraphics[width=8cm]{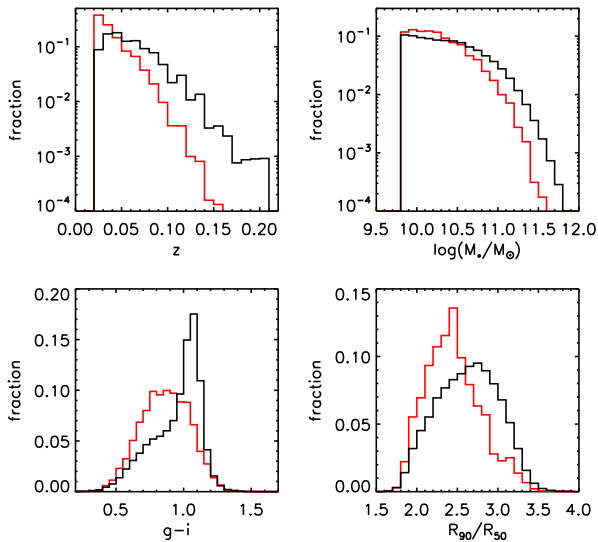}
 \caption{The normalized distributions of redshift, stellar mass, K-corrected g-i colour and $r$-band concentration index (defined as the ratio of the radii enclosing 90\% and 50\% of the total $r$-band light). The black and red lines are for samples S1 and S2, respectively.\label{fig:sample_dist}}
\end{figure}

We weight each galaxy by the inverse of $V_{max}$, which is defined as the total volume within which the galaxy could be located and make it into our sample. Because S1 is a sample selected by stellar mass, while S2 is a sample selected by stellar mass and 22 micron flux, we use different weightings for S1 and S2. Figure \ref{fig:sample_dist} shows the distribution of a number of different galaxy properties for galaxies in samples S1 and S2. Comparing with S1, S2 galaxies tend to have lower stellar masses, bluer g-i colours and lower concentrations. As we will show in the next section, the differences arise because passive galaxies are generally not detected at 22 micron. We note that the shape of the stellar mass distribution of S1 galaxies is consistent with the stellar mass function calculated by \citet{2003ApJS..149..289B} from 2MASS/SDSS data.

\subsection{Optical classification}
\label{ssec:data_optclass}

Here we use the classical \nii/\ha\ versus \oiii/\hb\ diagnostics \citep{1981PASP...93....5B, 1987ApJS...63..295V, 2003MNRAS.346.1055K, 2006MNRAS.372..961K} to classify the galaxies as star-forming or AGN from their optical emission line measurements. The fluxes of the key emission lines (\ha, \hb, \oiii$\lambda$5007, \nii$\lambda$6584 and \sii$\lambda\lambda$6717,31) are directly taken from the MPA/JHU catalogue. However, the errors on the emission line measurements from the pipeline are usually underestimated. We follow the recommendations on the webpage of MPA/JHU catalogue\footnote{http://www.mpa-garching.mpg.de/SDSS/DR7/raw\_data.html} and multiply the errors by 2.473, 1.882, 1.566, 2.039 and 1.621, respectively. To ensure the classifications are reliable, only the emission lines with S/N~$\ge$~3 are considered. In S1 there are 27755 sources above the line (K03 line) that \citet{2003MNRAS.346.1055K} suggest to separate AGNs from star-forming galaxies. We call them ``optical AGNs'' hereafter.

Following \citet{2004ApJ...613..109H}, we use the \oiii\ line luminosity as an indicator of AGN activity/black hole accretion rate. We use the Balmer decrement to correct the \oiii\ line for dust extinction, adopting the reddening curve in \citet{2007MNRAS.381..543W} and an intrinsic \ha/\hb\ ratio of 2.87 for star-forming galaxies and 3.1 for AGN \citep{1989agna.book.....O}. 67.6\% (18751) of the optical AGNs fall in the region between the K03 line and the maximum starburst line suggested by \citet{2001ApJ...556..121K}. A non-negligible fraction (between 10\% and 50\%)  of their \oiii\ luminosities can be contributed by star formation in the host galaxy \citep{2009MNRAS.397..135K}. In order to estimate the \oiii\ luminosity from the narrow-line region, we use the simple method suggested by \citet{2009MNRAS.397..135K} to separate the total \oiii\ luminosity into AGN and star formation components. The fraction of AGN contribution is calculated based on the galaxy's position on the BPT diagram (see their Figure 3).

In this work, ``strong'' AGNs are defined as optically-identified AGNs with \oiii\ luminosities larger than $10^7\mathrm{L_\odot}$. \citet{2003MNRAS.346.1055K} show that false BPT classification because of dilution by emission from HII regions in the surrounding host galaxy falling within the SDSS fibre aperture is not important for AGNs with \oiii\ luminosities greater than this value\footnote{We carried out test similar to what \citet{2003MNRAS.346.1055K} did by calculating the fraction of strong AGNs in narrow bins of stellar mass and redshift, and our results are similar to theirs.}.

We also classify the AGNs into Seyfert galaxies and LINERs (low ionization nuclear emission-line regions), according to the \sii/\ha\ ratio using Function 7 in the paper by \citet{2006MNRAS.372..961K}. A small fraction of AGNs are classified neither as Seyferts nor as LINERs, simply due to low signal-to-noise ratio of the \sii\ lines.

We select a sample of non-AGN hosts (``non-AGN''), which are either star forming galaxies (``SF'') or galaxies without detected emission lines (S/N $<2$, ``passive''). Passive galaxies are also required to have large concentration index $R_{90}/R_{50}>2.6$, high stellar mass surface density $\log(\mu_*/(\mathrm{M_\odot}/\mathrm{kpc^2}))>8.5$ and large 4000 \AA\ break strength $D_n\mathrm{(4000)}>1.6$. These values are the points where sharp transitions from young star-forming galaxies to old passive galaxies are observed to occur \citep{2003MNRAS.341...54K}.

In table \ref{tab:sample} we list the source numbers of each galaxy type. Because of the 22 micron luminosity cut, the fraction of passive galaxies in sample S2 is much smaller than in sample S1. We note a large fraction of objects (115512, 53.4\% of S1, ``ambiguous'') are not classified into any of the subclasses described above due to our strict criterion. We do not use them in our further analysis.

\begin{table}
 \centering
 \begin{tabular}{l|rrr}
   Opt-class & \multicolumn{1}{c}{S1} & \multicolumn{1}{c}{S2} \\ \hline
   All & 216272 (100.00\%) & 21942 (100.00\%) \\
   AGN & 27755 (12.83\%) & 8133 (37.07\%) \\
   Strong AGN & 7613 (3.52\%) & 4105 (18.71\%) \\
   Seyfert & 9776 (4.52\%) & 4171 (19.01\%) \\
   LINER & 16377 (7.57\%) & 3873 (17.65\%) \\
   SF & 23604 (10.91\%) & 11077 (50.48\%) \\
   Passive & 49401 (22.84\%) & 9 (0.04\%) \\
   Non-AGN & 73005 (33.76\%) & 11086 (50.52\%) \\
   Ambiguous & 115512 (53.41\%) & 2723 (12.41\%) \\ \hline
 \end{tabular}
 \caption{Sample S1 and S2. The first column shows the names of optical classifications. The numbers and the fractions are listed. Please see the text for the details of the sample definition and the classification criterion.\label{tab:sample}}
\end{table}

\subsection{SDSS quasars}
\label{ssec:data_quasar}

In our S1 and S2 samples, only narrow-line AGNs are included. We have extracted a sample of low redshift ($z<0.7$) type 1 AGN from the SDSS DR7 quasar catalogue \citep{2011ApJS..194...45S,2010AJ....139.2360S}. The upper redshift limit is chosen to ensure that the \oiii\ line still falls in the SDSS spectrum. We match this sample to the \wise\ catalogue within a 3" matching radius. Our type 1 sample includes 3165 quasars, of which 3086 (97.5\%) are detected in all \wise\ bands. Since the quasars are usually core-dominated, we use \wise\ magnitudes based on PSF-fit photometry. The \oiii\ line is corrected for extinction using the Balmer decrement measured from the narrow components of the \ha\ and \hb\ lines. We only use the emission lines with signal-to-noise ratio larger than 3. There are totally 3011 quasars with reliable \oiii\ fluxes and detected in all \wise\ bands, but only 592 of them have reliable Balmer decrement measurements, mostly because \ha\ line is in SDSS spectrum range only at $z\lesssim0.4$. For the other 2419 quasars, we apply the extinction correction assuming a typical Balmer decrement (\ha/\hb$\approx4.31$), which is estimated by taking the median value of the 592 quasars. We do K-correction to the \wise\ luminosities and colours of the quasars, using the QSO1 template from \citet{2007ApJ...663...81P}. The K-correction is small, less than 0.05 dex for 4.6 micron and less than 0.01 dex for 22 micron, at $z\approx0.7$. Different libraries \citep[e.g.][]{2010ApJ...713..970A} may give slightly different K-correction values, but as we will show later, compared to the observed scatter, the uncertainty of K-correction is negligible.

Throughout the paper, we adopt the concordance 0.7-0.3-0.7 cosmology \citep{2004PhRvD..69j3501T}.

\section{Mid-IR colours of local galaxies}
\label{sec:sdss}

\subsection{Stellar emission}
\label{ssec:stellar}

The \wise\ 3.4 and 4.6 micron bands are very similar to the \spitzer\ IRAC 3.6 and 4.5 micron bands. Previous studies show that in galaxies, the emission in this wavelength range is dominated by older stars, and luminosities are thus strongly correlated with stellar mass \citep{2012ApJ...744...17M,2012AJ....143..139E,2012ApJ...758...25H}.

\begin{figure}
 \centering
 \includegraphics[width=8cm]{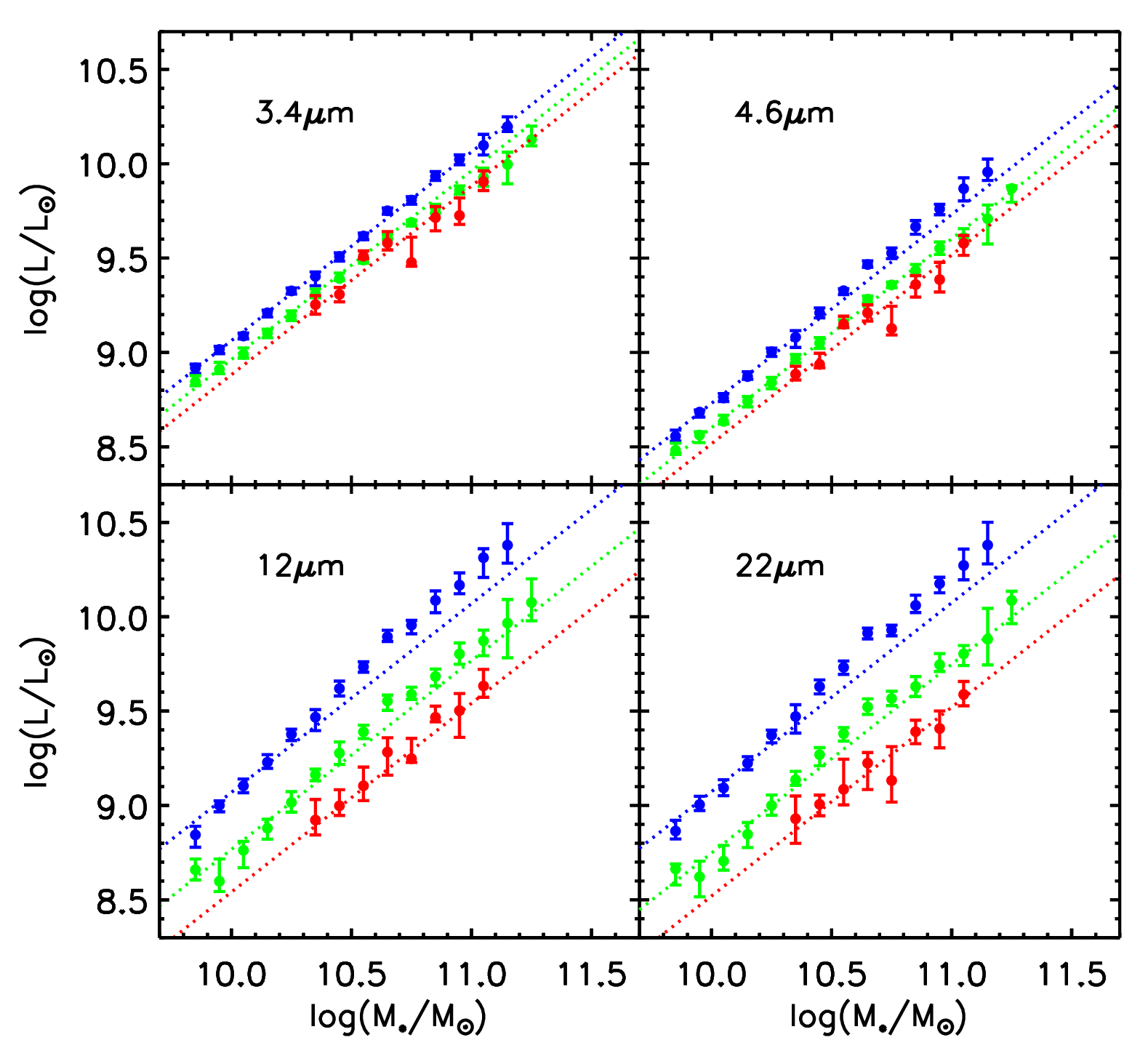}
 \caption{The \wise\ monochromatic luminosity as a function of stellar mass for S2 non-AGN galaxies. The blue, green and red points are for different 4000 \AA\ break strengths: $1.0<D_n\mathrm{(4000)}<1.3$, $1.3<D_n\mathrm{(4000)}<1.5$, $1.5<D_n\mathrm{(4000)}<1.8$. Any data bin with source number lower than 20 is dropped. The error bar shows the error on the median value at a confidence level of 95\%, estimated by bootstrapping within each data bin. The dotted lines are the best linear fits to the data, assuming a slope of 1.\label{fig:mass_lum}}
\end{figure}

We plot the $L-M_*$ relations of S2 inactive galaxies in Figure \ref{fig:mass_lum}. The four panels show luminosities evaluated at 3.4, 4.6, 12 and 22 micron. In order to investigate the sensitivity of \wise\ luminosites to star formation, we split our sample into 3 different bins of 4000 \AA\ break strength: $1.0<D_n\mathrm{(4000)}<1.3$; $1.3<D_n\mathrm{(4000)}<1.5$; $1.5<D_n\mathrm{(4000)}<1.8$. The 4000 \AA\ break strength, $D_n\mathrm{(4000)}$, can be regarded as an indicator of the specific star formation rate of the galaxy averaged over a timescale of around a Gyr. Unlike emission line fluxes (such as \ha), the 4000 \AA\ break is insensitive to extinction and to ``contamination'' from AGN. It allows a direct comparison of the stellar populations of AGN hosts and inactive galaxies \citep{2003MNRAS.341...33K}. As can be seen from Figure \ref{fig:mass_lum}, all the \wise\ luminosities have clear positive correlation with stellar mass, with slopes close to 1. With increasing wavelength, the difference between young and old galaxies becomes larger. This confirms that the 3.4 and 4.6 micron bands are indeed dominated by the light from old stars. Younger galaxies have lower mass-to-light ratios, probably due to additional contribution from thermally pulsing asymptotic giant branch (TP-AGB) stars. At 12 and 22 microns, the luminosity difference between galaxies of the same stellar mass with different 4000 \AA\ break strengths becomes very large. In these bands the emission from old stars is no longer dominant. Instead, the dust emission makes a substantial contribution to the total flux.

\begin{figure}
 \centering
 \includegraphics[width=8cm]{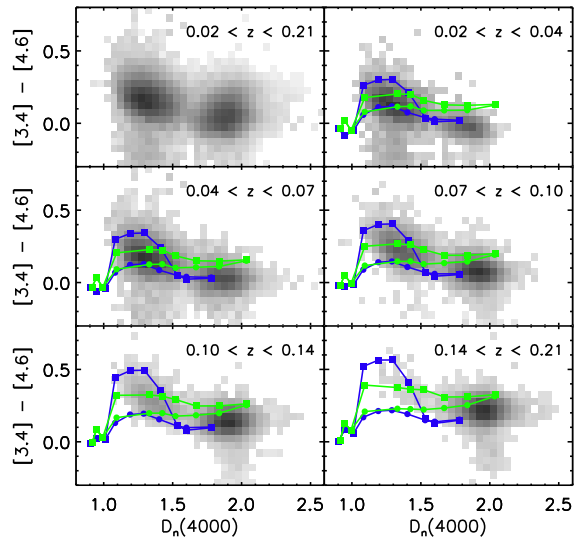}
 \caption{The distribution of S1 non-AGN galaxies is displayed as grey scaled histogram background on the [3.4] - [4.6] colour versus 4000 \AA\ break plane. Single stellar population models are overploted as blue (sub-solar, Z = 0.008) and green (solar metallicities, Z = 0.02) curves. The circles and squares are from BC03 and CB07 respectively. The data points along the curve are from the templates with stellar age of 0.005, 0.025, 0.1, 0.29, 0.64, 0.9, 1.4, 2.5, 5 and 11 Gyr. The model curves for each redshift bin are from the models convolved with \wise\ band filters at redshifts 0.03, 0.05, 0.08, 0.12 and 0.17, respectively.\label{fig:stellar_col}}
\end{figure}

In Figure \ref{fig:stellar_col}, we show the distribution of S1 non-AGN galaxies on the [3.4] - [4.6] colour versus 4000 \AA\ break (i.e. stellar age) plane. Since our aim in this section is to investigate the emission from {\em stars}, we make use of the S1 sample, which is not biased against galaxies with no ongoing star formation. Rather than K-correcting the colours, we show results in 5 narrow redshift ``slices''. The non-AGN sources show a clear bimodal distribution on the colour-age plane, reflecting the star-forming and passive populations of nearby galaxies. On average, the young star-forming galaxies are redder than older galaxies. 

We compare our data to the stellar population synthesis models of \citet{2003MNRAS.344.1000B} (BC03, circles) and an updated version of these model (CB07, squares). The major difference between the two models is the treatment of TP-AGB stars. In CB07 the dusty TP-AGB emission is more important and the models thus predict much redder IR colours at intermediate ages. In the plot, we show the predicted location of a ``simple stellar population'' (SSP) at a range of different ages, and for two different metallicities (see figure caption for details). The CB07 SSPs are clearly redder than the BC03 SSPs at stellar ages between $\sim$200 Myr and $\sim$2 Gyr, forming a red ``bump'' on the low metallicity CB07 curve. The [3.4] - [4.6] colours of the galaxies with the lowest 4000 \AA\ break strengths are well matched to the lower-metallicity single stellar population colours from the CB07 models, but not from BC03. This implies that the [3.4] - [4.6] colour difference between young and old galaxies is mainly due to emisson from TP-AGB stars. More detailed fits to models with more realistic star formation histories are required to draw more quantitative conclusions. We defer this to future work.

None of the model curves predict [3.4] - [4.6] colours redder than $\sim0.7$ and there appear to be very few normal galaxies with colours redder than this value in the data. In the next section, we analyze the mid-IR colours of AGN host galaxies.

\subsection{AGN host galaxies}
\label{ssec:agn_colour}

\begin{figure}
 \centering
 \includegraphics[width=8cm]{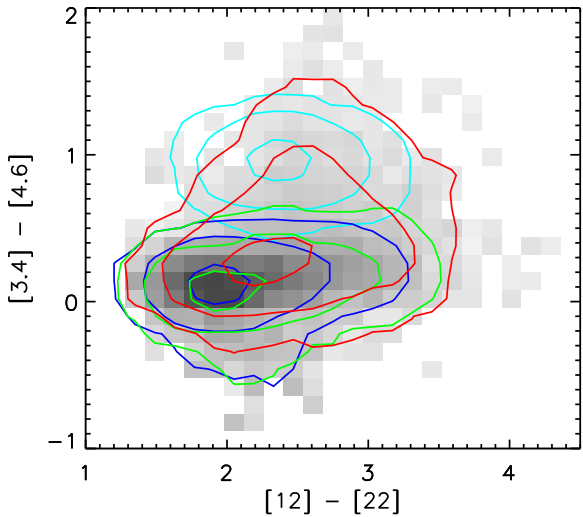}
 \caption{The \wise\ colour-colour diagram. The distribution of S2 galaxies is plotted as grey-scale background. The blue, green and red contours indicate the distributions of SF galaxies, weak AGNs and strong AGNs, respectively. The cyan contour is for the SDSS quasars.\label{fig:wcol}}
\end{figure}

Figure \ref{fig:wcol} displays the distribution of S2 galaxies in the [3.4] - [4.6] versus [12] - [22] \wise\ colour-colour plane. As can be seen, there is a clear peak at [3.4] - [4.6] colours near zero and [12] - [22] colours around 2. We have indicated the locations of different galaxy sub-populations, as well as quasars, using coloured contours as described in the figure caption. Weak AGNs have very similar colour distribution as star-forming galaxies. The peak of the colour distribution of strong AGNs is shifted with respect to that of star-forming galaxies, but the overlap between the two populations is very substantial. Only the SDSS quasars have \wise\ colours that are clearly disjoint from those of normal galaxies. As the host galaxy contamination is small for these systems, quasar colours reflect a ``pure'' AGN SED profile. We thus conclude that the [3.4] - [4.6] colours of most nearby type 2 AGNs are strongly affected by emission from stars. Only a minority of the strong type 2 AGNs have [3.4] - [4.6] and [12] - [22] colours that are similar to quasars.

We now examine how the \wise\ colours of AGNs vary as a function of 4000 \AA\ break strength (i.e. amount of recent star formation in the host galaxy) and as a function of our optically-defined accretion rate indicator based on the extinction-corrected \oiii\ line luminosity. We adopt the $M_\mathrm{BH}-\sigma$ relation from \citet{2009ApJ...698..198G} to calculate the BH mass:

\begin{equation}
 \log(\frac{M}{\mathrm{M_\odot}}) = (8.12\pm0.08) + (4.24\pm0.41)\log(\frac{\sigma}{200\,\mathrm{km\,s^{-1}}}).
\end{equation}

The velocity dispersions in the MPA/JHU catalogue are estimated by fitting the absorption lines in SDSS fiber spectra using a set of template spectra\footnote{http://spectro.princeton.edu/}. The statistical uncertainty is around 10 km/s, leading to $\sim0.05-0.2$ dex uncertainty of black hole mass, smaller than the intrinsic scatter of the $M_\mathrm{BH}-\sigma$ relation \citep[e.g.][0.44 dex]{2009ApJ...698..198G}. Due to SDSS spectral resolution of $\sim70$ km/s, the black hole mass estimation is no longer reliable at $\lesssim10^{6.2}\,\mathrm{M_\odot}$. Only a very small fraction of our objects are in this range and they do not affect our results. A large fraction of our AGNs are hosted by late type galaxies (see also Section \ref{ssec:opt-ir}), for which the disk component is not negligible within the fiber aperture. In this paper we do not perform relevant corrections, e.g. by bulge-disk decomposition, to the derived velocity dispersions. We note, however, this does not significantly affect our results. Throughout this paper, we will use the Eddington parameter $L$\oiii$/M_\mathrm{BH}$ as the optical indicator of the central BH accretion activity level.

\begin{figure}
 \centering
 \includegraphics[width=8cm]{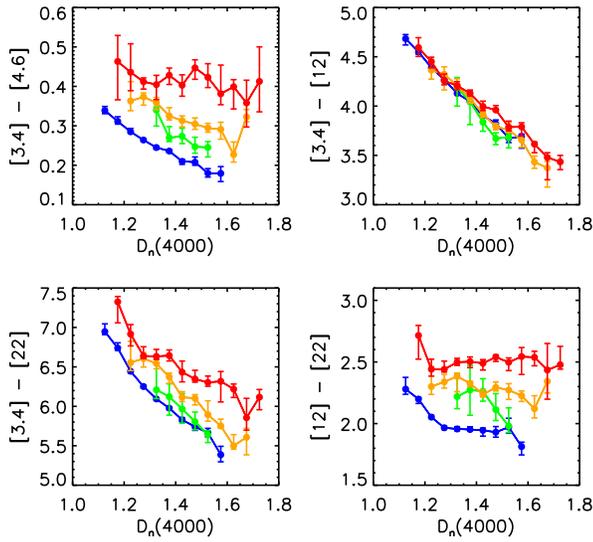}
 \caption{The \wise\ colours as function of 4000 \AA\ break. The plots are for sample S2 galaxies with $0.07<z<0.11$. Blue curves are for normal star-forming galaxies. The green, orange and red lines are for strong AGNs in different Eddington ratio bins: $\log(L$\oiii$/M_\mathrm{BH})<-0.56$, $-0.56<\log(L$\oiii$/M_\mathrm{BH})<0.01$, $\log(L$\oiii$/M_\mathrm{BH})>0.01$, respectively (all the values are in solar units). The error bar is the 1-$\sigma$ error on the median value, catculated from bootstrap resampling each data bin. Any data bin with source number less than 20 is discarded.\label{fig:d4000_col}}
\end{figure}

Figure \ref{fig:d4000_col} shows the four different \wise\ colours of strong AGNs as function of 4000 \AA\ break (i.e. host galaxy sSFR). Results are shown for AGN divided into 3 different ranges in $L$\oiii/$M_\mathrm{BH}$ (see figure caption). Results for star-forming galaxies are also plotted in blue for comparison. In order to minimize the scatter due to the fact that our colours have not been K-corrected, we only use the galaxies within a relatively small redshift range $0.07<z<0.11$. The redshift cut we use includes galaxies in the stellar mass range $10.4<\log(M_*/\mathrm{M_\odot})<11.6$ (see Figure \ref{fig:zmass}). Our conclusions do not change if the analysis is done in other redshift and stellar mass ranges.

\wise\ colours are redder at higher sSFR. The [3.4]-[12] \wise\ colour is most sensitive to star formation. There is no difference between AGNs and star-forming galaxies in the top right panel of Figure \ref{fig:d4000_col}, indicating that the [3.4]-[12] colour is insensitive to AGN activity. In other \wise\ colours, the AGN contribution is more prominent and stronger AGNs are redder at any fixed $D_n\mathrm{(4000)}$. The [3.4]-[4.6] and [12]-[22] colours are most sensitive to accretion rate and least sensitive to star formation. This is consistent with the observed peak shift between strong AGNs, weak AGNs and star-forming galaxies shown in Figure \ref{fig:wcol}. 

In the following sections, we drop the 12 and 22 micron bands from our analysis, and focus on the 4.6 micron luminosity as our main indicator of AGN activity at mid-IR wavelengths. Although we see in Figure \ref{fig:d4000_col} that the difference in colour between different $L$\oiii$/M_\mathrm{BH}$ bins is in fact {\em larger} for [12] - [22] than for [3.4] - [4.6], the loss in sample size incurred by requiring our AGN samples to be complete at 22 microns is too large. All the plots in the rest of the paper are based on sample S1.

\section{Mid-IR properties of local AGNs}
\label{sec:optagn}

In this section, we will use the 4.6 micron luminosity, corrected for the contribution from stars, as a way to parametrize the IR properties of the AGN in our sample. We will then compute how the total IR emissivity from AGN is distributed among galaxies with different masses, structural parameters and stellar populations. Finally we will compare these distribution functions with similar ones computed for the total \oiii\ emissivity.

\subsection{AGN IR luminosity}
\label{ssec:lir}

In the previous section, we showed that the [3.4] - [4.6] colour is the best \wise\ indicator of AGN activity. This implies that the 4.6 micron luminosity can be used as an estimate of the mid-IR luminosity that originates from the torus itself, provided that we are able to subtract the contribution that originates from stars. In this section, we devise a method for performing this subtraction.

Because the S1 sample is large, we find non-AGN control galaxies with properties that closely match those of the AGN host galaxies. We match the AGNs to non-AGN galaxies in stellar mass, 4000 \AA\ break, redshift and stellar mass surface density. Matching in stellar mass and $D_n\mathrm{(4000)}$ ensures that the control galaxies have the same total stellar masses and stellar population ages as the AGN hosts. Matching in both redshift and stellar surface mass density ensures that the 3" diameter SDSS fiber aperture samples the same fraction of the total light from the hosts in the two samples. In order to minimize the scatter, we use strict matching criteria: stellar mass within $\pm0.01$ dex, 4000 \AA\ break within $\pm 0.025$, redshift within $\pm0.02$, and stellar mass surface density within $\pm0.2$ dex. For every AGN we estimate the contribution of stars to the total 4.6 micron luminosity from the median 4.6 micron luminosity of all the matched control galaxies. Typically there are 8 non-AGN ``neighbours'' for each of AGN to give reasonable estimation of the non-AGN component. We then subtract this estimate of the non-AGN component from the observed 4.6 micron luminosity to get the ``pure'' AGN 4.6 micron luminosity $L_{4.6\mathrm{\mu m,AGN}}$. This correction should be regarded as a statistical one. We do K-correction to this ``pure'' AGN component with the QSO1 template from \citet{2007ApJ...663...81P}. In our redshift range, the correction is less than 5\% due to flat quasar SED in this wavelength.

\begin{figure}
 \centering
 \includegraphics[width=8cm]{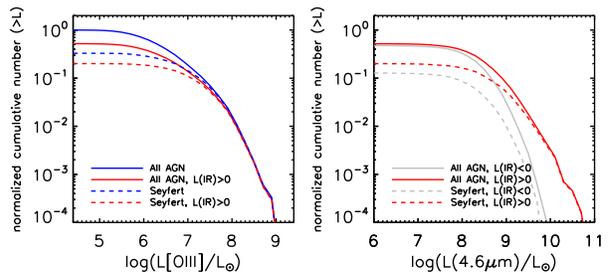}
 \caption{The number of objects as a function of \oiii\ luminosity (left panel) and 4.6 micron luminosity (after host correction, right panel), normalized by the total number of S1 AGNs. The solid and dashed lines are for all optically selected AGNs and Seyfert galaxies, respectively. In left panel, the blue curve is for all the objects in the subsample and the red curve is for the objects with positive 4.6 micron luminosities. In right panel, the red curve is the same population as left panel. The grey curve is the AGNs with negative 4.6 micron luminosities, binned by the absolute values of their luminosities. The error bars are estimated by bootstrapping within the whole S1 AGN sample.\label{fig:ir_numdist}}
\end{figure}

In some cases, particularly when the AGN luminosity is low, the resulting flux will be negative. Figure \ref{fig:ir_numdist} shows that a substantial fraction of objects have negative fluxes when $L$\oiii$\lesssim10^7\,\mathrm{L_\odot}$ or $L_{4.6\mathrm{\mu m,AGN}}\lesssim3\times10^8\,\mathrm{L_\odot}$. We also estimate the uncertainty of individual object by calculating the scatter of the 4.6 micron luminosities of its ``neighbours'' used for host subtraction. This yields typical uncertainty value of $\sim2-7\times10^8\,\mathrm{L_\odot}$. Below this luminosity level, the nuclear emission is poorly determined due to host contamination. Comparing with the whole AGN sample, the Seyfert galaxies are less affected due to their higher nuclear luminosities. Seyferts with positive fluxes always account for a larger proportion even when summing up to the lowest luminosities bins. The 4.6 micron luminosities of Seyferts are much better recovered individually, allowing a direct comparison between IR and \oiii\ luminosities. We note that the exact definition of AGNs does not affect our results because we have subtracted the host star formation contribution to both the \oiii\ and IR luminosities.

\begin{figure}
  \centering
  \includegraphics[width=8cm]{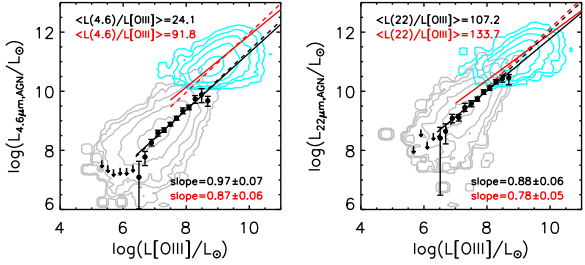}
  \caption{Left: 4.6 micron AGN luminosity versus \oiii\ luminosity for S1 Seyferts. The grey and cyan contours are S1 Seyferts (with positive AGN 4.6 micron luminosity) and SDSS quasars respectively. The black dots and upper limits show results of stacked S1 Seyferts. The black solid line is a linear fit to the black dots and the resulting slope is shown in bottom-right corner. The black dashed line is a linear fit assuming the 4.6 micron luminosity is proportional to \oiii\ luminosity. The median value of the IR-to-\oiii\ ratio is shown on top-left corner. The red lines are similar to black lines but for SDSS quasars. Right: similar to the left panel, except that we use 22 micron luminosities and the local Seyferts are from S2 sample.}
  \label{fig:lo3lir}
\end{figure}

The left panel of Figure \ref{fig:lo3lir} shows that there is a good correlation between the corrected 4.6 micron monochromatic ($\nu L_\nu$) ``pure'' AGN luminosity and the \oiii\ luminosity for Seyferts, though the scatter is as large as $\sim0.28$ dex on average. We do not include the LINERs in the plot, because the 4.6 micron flux from the central source cannot be estimated accurately due to host galaxy contamination. We stack the Seyferts in different \oiii\ luminosity bins to reduce the uncertainty (black dots and upper limits). We limit the linear fitting to the objects with $L$\oiii$>3\times10^6\mathrm{L_\odot}$ (66\% of the Seyferts in volume-weighted number), because at lower luminosity level it is too noisy to recover nuclear emission even with the stacking technique. The linear fit to the bright narrow line Seyferts gives a correlation of $L_{4.6\mathrm{\mu m,AGN}} \propto L$\oiii$^{0.96\pm0.07}$ (solid line). If we assume the IR luminosity is proportional to \oiii\ luminosity, then we get a median IR-to-\oiii\ luminosity ratio of $\sim24$ (dashed line). We note these values are only valid for bright objects. We also compare the results with SDSS quasars. The quasars extend the correlation a further 2 orders of magnitude in \oiii\ and in IR luminosity, although it seems there is a systematic offset between quasars and the Seyferts.

Interestingly, if we plot the AGN IR luminosities at 22 micron, estimated with similar host subtraction methods, the Seyfert-quasar offset becomes much smaller (see right panel of Figure \ref{fig:lo3lir}). We thus hypothesize that offset may be caused by the intrinsic obscuration of torus: longer wavelengths are less absorbed, and type 1 AGNs may be systematically less obscured than type 2 AGNs\footnote{We note that we do not find correlation between the [4.6] - [12] or [4.6] - [22] colors of the AGN component and \oiii\ luminosity within Seyfert sample or within the quasar sample.}.

\subsection{Comparison of the distribution of total IR luminosity and total \oiii\ luminosity from AGN as a function of host galaxy properties}
\label{ssec:opt-ir}

\begin{figure*}
 \centering
 \includegraphics[width=16cm]{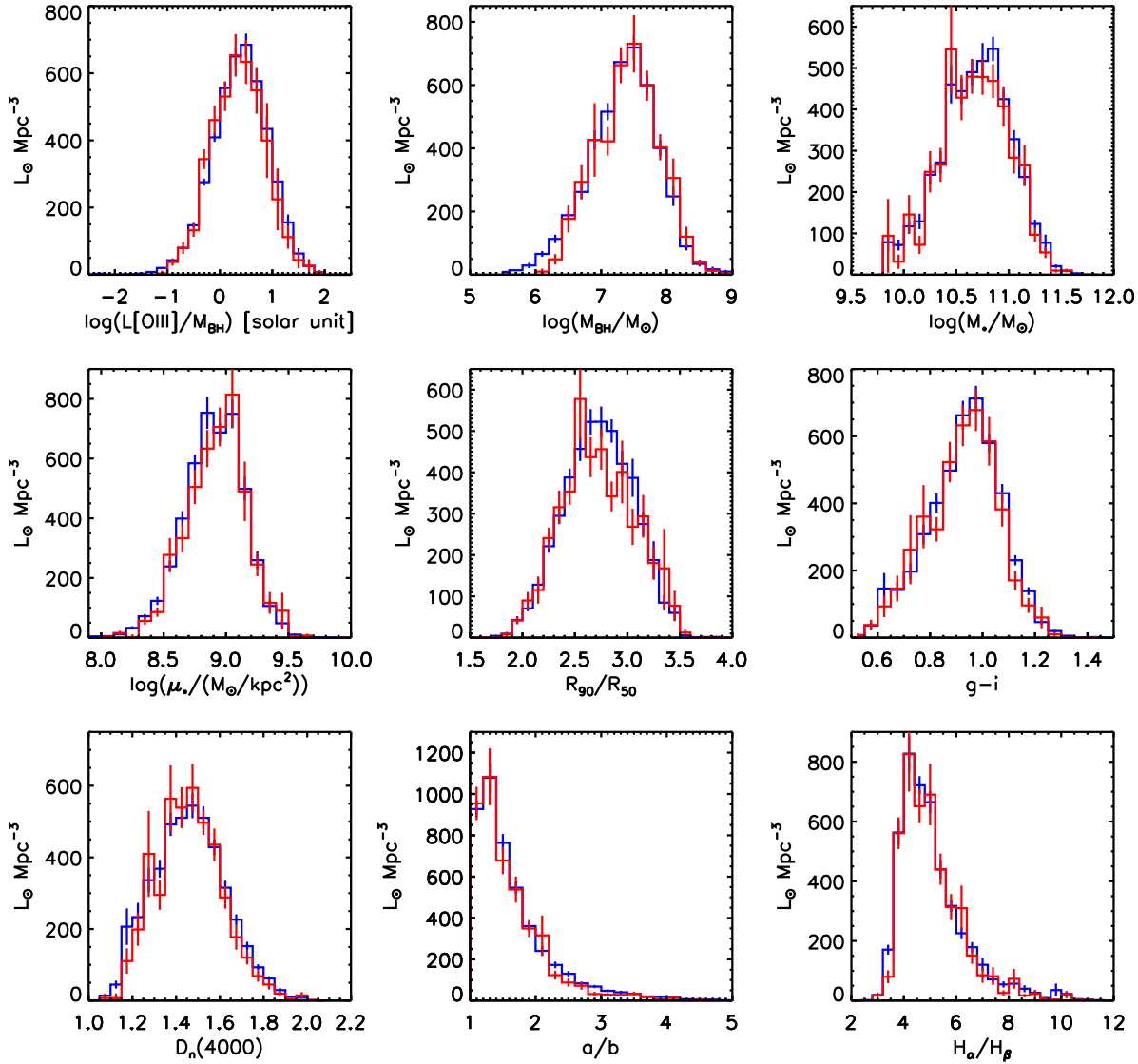}
 \caption{The total \oiii\ and IR emissivity as a function of various AGN properties for Seyferts (see text for details). The blue histogram is for the \oiii\ luminosity density and the red histogram is for the 4.6 micron luminosity density. The red histogram is scaled down by a factor of 24.2 to compensate the constant ratio between 4.6 micron and \oiii\ luminosities calibrated in Figure \ref{fig:lo3lir}.\label{fig:optir_dist_seyf}}
 \end{figure*}

\citet{2004ApJ...613..109H} investigated the {\em integrated} \oiii\ luminosity from type 2 AGN binned up as a function of stellar mass, of stellar surface mass density, of concentration index and of 4000 \AA\ break strength $D_n\mathrm{(4000)}$. They showed that most of the present-day accretion traced by the \oiii\ line is taking place in galaxies with young stellar ages ($D_n\mathrm{(4000)}<1.6$), intermediate stellar masses ($10^{10}$ - few $\times 10^{11}\,\mathrm{M_\odot}$), high surface mass densities ($3\times10^8 - 3\times10^9\,\mathrm{M_\odot/kpc^2}$), and intermediate concentrations ($R_{90}/R_{50} = 2.2 - 3.0$). In this section we carry out the same exercise using the integrated 4.6 micron luminosity and compare the results with what is obtained for the integrated \oiii\ luminosity. We note that we use 4.6 micron luminosities that are corrected for emission from stars and \oiii\ luminosities that are corrected for extinction and for the contribution from star formation for this exercise.

The results are shown in Figure \ref{fig:optir_dist_seyf}, where blue histograms show the distribution of the total \oiii\ emission from the Seyfert sample and the red histograms show the distribution of the total 4.6 micron luminosity as a function of a wide variety of different host galaxy parameters. From left to right, and from top to bottom, the host galaxy properties investigated in Figure \ref{fig:optir_dist_seyf} are the following:

\begin{enumerate}
  \item The \oiii\ line luminosity normalized by the black hole mass (Eddington parameter).
  \item The black hole mass estimated from the stellar velocity dispersion.
  \item Stellar mass $M_*$.
  \item Stellar mass surface density $\mu_*$.
  \item The concentration index, defined as the ratio between 90\% light radius and 50\% light radius ratio $R_{90}/R_{50}$.
  \item Rest-frame g-i colour.
  \item 4000 \AA\ break, $D_n\mathrm{(4000)}$.
  \item Galaxy inclination estimated from the ratio of the major-to-minor axes, $a/b$.
  \item The Balmer decrement calculated from the ratio of H$\alpha$ to H$\beta$ line fluxes.
\end{enumerate}

Comparison of the red and blue histograms in Figure \ref{fig:optir_dist_seyf} indicates that the host galaxies of the Seyferts producing the integrated 4.6 micron luminosity and the integrated \oiii\ are identical. This provides strong support for the standard Unified Model \citep{1995PASP..107..803U}. We again note, however, that in our sample, detection of low luminosity mid-IR emission from the central source is impossible because of host galaxy contamination. In order to avoid spurious result due to objects with negative IR luminosities, we check the emissivity distributions of a sample of 671 IR-bright Seyferts with posivite AGN 4.6 luminosities larger than $10^{9.5}\,L_\odot$\footnote{At this level, the AGN component is comparible or stronger than the host component at 4.6 micron. The individual detection of AGN component is relatively reliable.}.

\begin{figure}
  \centering
  \includegraphics[width=8cm]{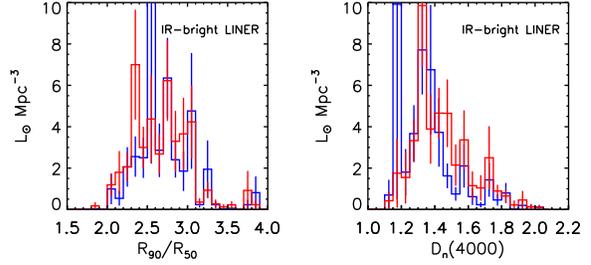}
  \caption{Similar to Figure \ref{fig:optir_dist_seyf} but for IR luminous LINERs. For simplicity, results are shown only for two of the parameters: concentration index $R_{90}/R_{50}$ and 4000 \AA\ break $D_n\mathrm{(4000)}$. Here the 4.6 micron emissivity histogram is scaled down by a factor of 155.9, which is larger than the factor 24.1 used in Figure \ref{fig:optir_dist_seyf}. This is simply because the IR selected subsample is biased towards higher IR-to-optical ratios.\label{fig:optir_dist_liner}}
\end{figure}

We also investigate a sample of 134 IR-bright LINERs, as shown in Figure \ref{fig:optir_dist_liner}. Despite the large uncertainties, the IR-bright LINERs show consistent \oiii\ and IR luminosity distributions. However, we fail to establish this kind of \oiii-IR link for all the LINERs. We find the \oiii\ luminosity distribution of the whole LINER sample shifts towards bulge-dominated galaxies with high 4000 \AA\ break strengths. It is probably due to the fact that LINERs hosted by old elliptical galaxies have relatively weaker AGN luminosities than Seyferts, and we are not able to detect the nuclear IR component for these low luminosity objects.

\section{Investigation of the sub-population of AGNs selected using mid-IR colours}
\label{sec:iragn}

In this section, we investigate the sub-population of galaxies identified as AGN from their WISE photometry, but not from optical emission line diagnostics. We begin with a brief review of IR-only AGN selection methods.

\subsection{\wise\ IR AGN selection}
\label{ssec:irselect}

The original IR colour-colour selection techniques were based on the \spitzer\ IRAC colours \citep{2004ApJS..154..166L,2005ApJ...631..163S,2006ApJS..166..470R,2008ApJ...687..111D}. The selection was then extended to \wise\ bands by synthesizing data from multiwavelength observations and template studies \citep{2010ApJ...713..970A}. Later, using real data, \citet{2012ApJ...753...30S} adopted a simpler criterion: $\mathrm{[3.4]-[4.6]}>0.8$. We note that their selection criterion was tuned to AGN searches at higher redshifts. If we adopt this cut, we only find 435 sources in sample S1.

\begin{figure}
 \centering
 \includegraphics[width=8cm]{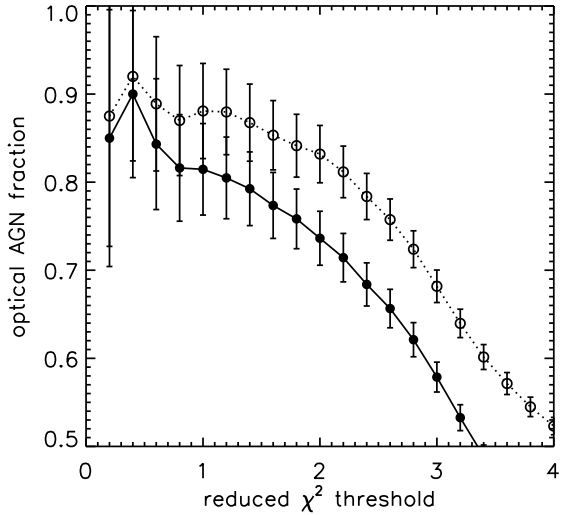}
 \caption{The fraction of optically-identified AGNs in the population that meet our mid-IR power-law selection criterion $\alpha<-0.5$ as a function of reduced $\chi^2$ threshold. The solid circles and empty circles indicate strong and all optical AGNs, respectively. The error bars are simple Poisson counting errors.\label{fig:irpow_chisq}}
\end{figure}

Another method is IR power-law selection, which was first applied to \spitzer\ IRAC data \citep{2006ApJ...640..167A} and is considered to be a stricter way to select AGN than colour-colour selection \citep{2008ApJ...687..111D}. It is based on the fact that the AGN continuum in quasars is known to have a power-law form. In practice, we make use of all four \wise\ bands and fit the broadband SED as follows:

\begin{equation}
 0.4 \times M_\mathrm{AB} = -\alpha \times \log(\lambda/1\,\mathrm{\mu m}) + c.
\end{equation}

Here $M_\mathrm{AB}$ is the monochromatic AB magnitude and $\lambda$ is the effective wavelength in each \wise\ bands. The free parameter $\alpha$ is spectral slope ($f_\nu\propto\nu^\alpha$). AGNs are required to have spectral slope that are sufficiently red ($\alpha<-0.5$). The quality of the fit, i.e. the similarity of the SED shape to a pure power law, can be quantified by $\chi^2$. A sample selected with looser $\chi^2$ threshold will include more galaxies, but also be contaminated by more star forming galaxies. In Figure \ref{fig:irpow_chisq}, we plot the fraction of optically-selected AGNs as a function of threshold in reduced $\chi^2$ statistics. Results are shown for all AGN (open symbols) and for strong AGN (filled symbols). As can be seen the fraction of optically-identified sources drops sharply above $\chi^2\approx1.5$, particularly for strong AGN. We therefore select this as a threshold, which yields a sample of 503 IR-selected power-law AGNs.

\begin{table}
  \centering
  \begin{tabular}{l|rrr}
    IR selection & \multicolumn{1}{c}{ALL} & \multicolumn{1}{c}{optical AGN} & \multicolumn{1}{c}{optical strong AGN} \\ \hline
    col & 435 & 379 (87.1\%) & 358 (82.3\%) \\
    pow & 503 & 434 (86.3\%) & 394 (78.3\%) \\
    col \& pow & 284 & 255 (89.8\%) & 243 (85.6\%) \\
    all & 654 & 558 (85.3\%) & 509 (77.8\%) \\ \hline
  \end{tabular}
  \caption{The number census of the IR selected AGN samples. ``col'' and ``pow'' are the [3.4] - [4.6] colour selection and the power law selection, respectively. Symbol ``\&'' means the sources are selected by both methods. ``all'' means the sources selected by any of the IR selections.\label{tab:iragn}}
\end{table}

Table \ref{tab:iragn} shows the number of the IR AGNs with different selection methods. In total, we find 654 IR AGN. The fraction of optically identified AGN is high (85.3\%). We note that this is a much smaller number than could be identified optically. It is clear that IR AGN selection methods based on \wise\ colours will miss a large fraction of type 2 AGNs at low redshift. It is not a surprising result. Previous works based on AGN samples selected in other bands have shown that this is the same situation at higher redshifts \citep{2008ApJ...680..130C,2009A&A...507.1277B,2012arXiv1209.6055A}. Simple mid-IR colour selected AGN samples are clearly biased. It is necessary to use decomposition methods in mid-IR regime, like we have done in previous section \citep[see also, e.g.][]{2011MNRAS.414.1082M}, for an unbiased AGN study.

\subsection{The SEDs of AGN selected at mid-IR wavelengths}
\label{ssec:iragn_sed}

\begin{figure}
 \centering
 \includegraphics[width=8cm]{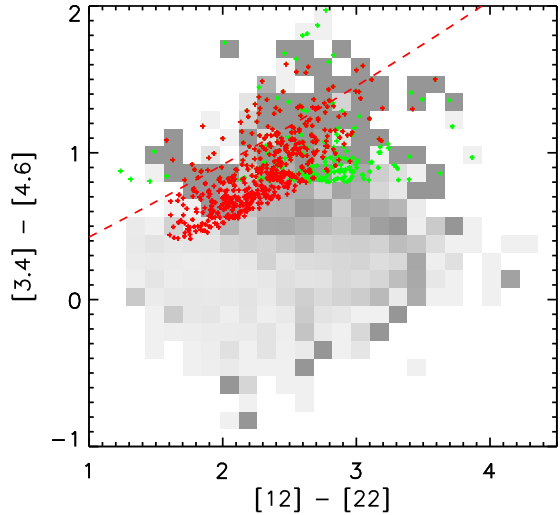}
 \caption{Similar to Figure \ref{fig:wcol} but the grey-scale background shows the location of strong optical AGNs. The green and red crosses indicate [3.4] - [4.6] colour and power-law selected AGNs. The red dashed line is the track of a pure power-law spectrum.\label{fig:wcol_iragn}}
\end{figure}

In Figure \ref{fig:wcol_iragn}, we show the distributions of the IR AGN samples on the \wise\ colour-colour diagram as in Figure \ref{fig:wcol}. Green crosses indicate sources selected by the simple [3.4] - [4.6] colour cut, while red crosses indicate power-law sources. The colours of IR AGNs are consistent with typical quasar SEDs and clearly avoid the locus of star-forming galaxies.

\begin{figure}
 \centering
 \includegraphics[width=8cm]{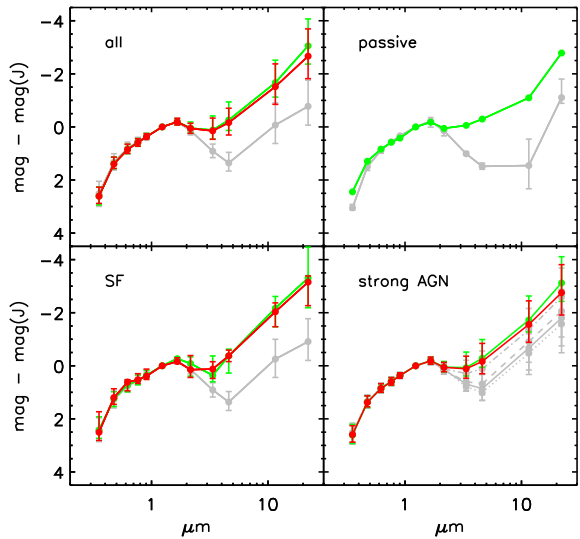}
 \caption{Rest frame median SEDs of different subsamples of S2 galaxies. All SEDs are normalized at $\sim$1 micron (J band). Grey, green and red colours are for the whole S1 sample, IR colour selected AGN and power law selected AGN, respectively. In the bottom-right panel, we also split all the S1 strong AGNs with detected 4.6 micron AGN luminosity into three luminosity bins, $L(4.6\,\mathrm{\mu m,AGN})<10^{8.8}\,\mathrm{L_\odot}$ (dotted line), $10^{8.8}\,\mathrm{L_\odot}<L(4.6\,\mathrm{\mu m,AGN})<10^{9.3}\,\mathrm{L_\odot}$ (dashed line) and $L(4.6\,\mathrm{\mu m,AGN})>10^{9.3}\,\mathrm{L_\odot}$ (dash-dotted line). The error bar is the 1$\sigma$ scatter within the bin. The four panels show SEDs for galaxy subsets with different optical classifications.\label{fig:iragn_sed}}
\end{figure}

Our photometric data covers 5 SDSS bands, 3 near-IR bands (2MASS bands, extracted from NYU Value-Added Galaxy Catalog produced by \citet{2005AJ....129.2562B}) and 4 \wise\ bands. We use these bands to build a SED for each source in S1. We calculate the AB magnitudes in the rest frame, interpolated from neighbouring data points. The detection rate of \wise\ 12 and 22 micron bands is relatively low in sample S1, so here we use sample S2 instead. We note that if we use S1 we get similar result.

Figure \ref{fig:iragn_sed} shows that both IR selection methods lead to similar SED shapes. IR AGNs are similar to field galaxies in the optical and the near-IR, but clearly different beyond 3 micron. All galaxies, except the passive ones which have little dust emission, show a clear turnover at $\sim$5 micron. This is the point where the dust emission starts to dominate the total radiation output for galaxies with ongoing star formation or nuclear activity. In most cases the AGN component is not prominent in mid-IR. Only the AGNs with the highest nuclear IR luminosities ($L(4.6\,\mathrm{\mu m,AGN})>10^{9.3}\,\mathrm{L_\odot}$, dash-dotted line) can be distinguished by their mid-IR colours when the turnover moves to shorter wavelength ($\sim$2-3 micron). This strongly affects the [3.4] - [4.6] colours. We note there are 9 passive galaxies in S2, shown in top-right panel. The origin of their 22 micron fluxes is still unknown. One possible explanation is highly dust obscured star formation and/or AGN activity, which the optical emission line diagnostics may fail to identify \citep[see, e.g.][]{2007MNRAS.376..416R,2009ApJ...693..340B}. We defer this to future work.

\subsection{Optical properties of AGN selected at mid-IR wavelengths}
\label{ssec:iragn_prop}

\begin{figure}
 \centering
 \includegraphics[width=8cm]{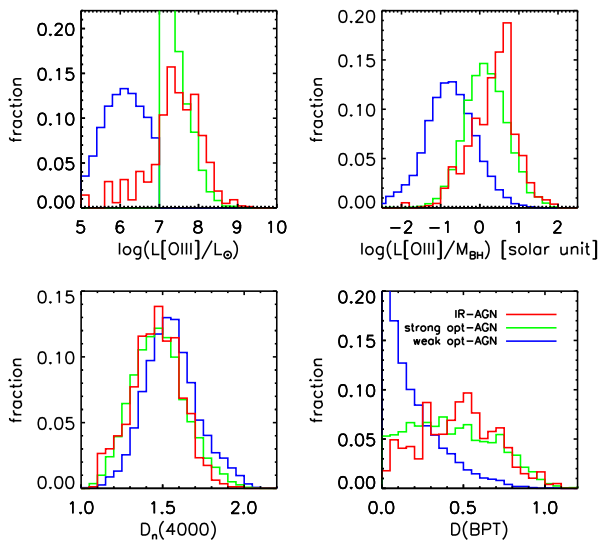}
 \caption{The number distributions of IR AGNs (red), strong optical AGNs (green) and weak optical AGNs (blue). The properties shown here are \oiii\ luminosity, Eddington parameter, 4000 \AA\ break and D(BPT). D(BPT) is the distance to the AGN/SF separation line defined by \citet{2003MNRAS.346.1055K} on the BPT diagram. ``Pure'' AGNs have the largest D values.\label{fig:iragn_dist}}
\end{figure}

Since both IR AGN selection methods lead to consistent SED shapes, we simply combine both IR AGN samples. In Figure \ref{fig:iragn_dist} we compare IR-selected AGN with optically-identified weak and strong AGN. The properties we investigate are \oiii\ luminosity, Eddington parameter, 4000 \AA\ break and D(BPT), the distance to the AGN/SF separation line defined by \citet{2003MNRAS.346.1055K} on BPT diagram (``pure'' AGNs have the largest D values). Unlike Figure \ref{fig:optir_dist_seyf}, this plot shows fraction by {\em number} rather than fraction of the integrated IR or \oiii\ emissivity. The main thing we learn is that AGNs selected at low redshifts using mid-IR colours have similar \oiii\ luminosities and 4000 \AA\ break strengths as strong optical AGN, but have values of the Eddington parameters and the D parameter that are slightly higher. This bias arises because the mid-IR colours are much more sensitive to star formation in the host galaxy than the \oiii/\hb\ and \nii/\ha\ ratios. As a result, only strongly accreting black holes with high Eddington parameters are selected by IR colour-based techniques.

\subsection{Optically-unidentified IR AGN}
\label{ssec:ironly}

Though the IR AGNs are found to be similar to strong optical AGNs, there is a small fraction ($\lesssim15\%$) of objects not identified as AGNs from their optical emission lines. This holds for both IR selection techniques. In total there are 96 (14.7\%) IR AGNs that are not classified as AGN in the optical. We call them ``IR-only'' AGNs for short.

There are two reasons why optical identification might fail. One is because they are mis-classified as star-forming systems. The optical classification fails because of emission-line contamination from the host galaxy. The second possibility is that at least one of the four emission lines required for reliable BPT classification is not detected.

We find 20\% of IR-only AGNs are optically identified as star-forming galaxies (hereafter we call them SF-IR-only AGNs). Most (15 out of 19 sources) are relatively metal rich ($\log($\nii/\ha$)>-0.6$) and are located at the border between the AGN and star-forming populations. There are also 4-metal poor SF-IR-only AGNs ($\log($\nii/\ha$)<-0.6$) that fall on the left side of the BPT diagram. Metal-poor AGNs are rare and occur in less massive galaxies \citep{2006MNRAS.371.1559G}. Three of these objects do not have red [3.4] - [4.6] or [12] - [22] colours, so it is difficult to judge whether these galaxies are true AGNs or not. Interestingly, one object has [3.4] - [4.6] $=1.1$ and [12] - [22] $=7.2$. It is clearly an AGN-dominated object with relatively low stellar mass $\log(M_*/\mathrm{M\odot})=9.83$.

The majority of our IR-only AGNs (77 objects, 80\% of the population), are optically unidentified, because one or more emission lines are not detected with sufficient signal-to-noise ratio. In most cases, it is the \hb\ line measurement that has low S/N. We can estimate a lower limit on their \oiii/\hb\ ratio that places them well into the Seyfert regime. In conclusion, as far as we can tell, IR-selected AGN without optical emission line classification {\em do not constitute a special class of object}.

\section{Discussion}
\label{sec:discussion}

In this paper, we have matched a large sample of SDSS galaxies with redshifts in the range $0.02<z<0.21$ with mid-IR photometry from the \wise\ survey. Our aim was to investigate the host galaxy properties of AGN by using the mid-IR luminosity as our AGN activity indicator, and then to compare the results to previous studies, which have used the \oiii\ line luminosity as the main diagnostic of AGN activity.

As an AGN activity indicator, the \oiii\ line luminosity has the advantage that it is relatively insensitive to contamination by ionized gas excited by young stars in the host galaxy (the \oiii\ luminosity produced in high metallicity HII regions is known to be weak). However, because the bulk of the \oiii\ emission excited by radiation from the accretion disk arises from gas located at distances of a few hundred parsecs from the galaxy center, the \oiii\ luminosity is a rather indirect indicator of current accretion onto the central black hole. In contrast, recent high-resolution observations indicate that the scale of the IR-emitting ``torus'' around the black hole might be no more than a few parsecs in radius \citep[e.g.,][]{2004Natur.429...47J,2005astro.ph.12025E,2007A&A...474..837T,2008A&A...486L..17B,2009ApJ...705L..53B,2009A&A...502...67T,2009A&A...493L..57K,2011A&A...536A..78K,2011A&A...527A.121K,2012ApJ...755..149H}. It thus provides a probe of accretion onto the black hole on much smaller scales. However, a significant fraction of the {\em total} mid-IR emission from galaxies arises from stars. In the 3.4 and 4.6 micron bands, emission from stars with ages greater than $\sim1$ Gyr dominates, and at longer wavelengths emission from the dusty interstellar medium becomes important.

In this paper, we use the 4000 \AA\ break strength, $D_n\mathrm{(4000)}$, as our main probe of present-to-past averaged star formation in galaxies. We first carry out a systematic study of how the mid-IR colours of AGN hosts vary as a function of both $D_n\mathrm{(4000)}$ and optical ``Eddington parameter'' ($L$\oiii$/M_\mathrm{BH}$), finding that the [3.4] - [4.6] micron colour to have the weakest dependence on $D_n\mathrm{(4000)}$, but strong dependence on $L$\oiii$/M_\mathrm{BH}$. We use a ``pair-matching'' technique introduced by \citet{2006MNRAS.367.1394K} to statistically subtract the 4.6 micron stellar emission contributed by the host galaxies of the the AGN in our sample, by extracting samples of non-AGN with similar redshifts, stellar masses, sizes and 4000 \AA\ break strengths as the AGN host galaxies. We use these corrected 4.6 micron luminosities to parametrize the strength of the central torus emission for the AGN in our sample. We show that intrinsic 4.6 micron AGN luminosities can be recovered for most Seyferts, but only statistically for LINERs.

By combining our sample of Seyferts with a sample of type 1 AGN and quasars at $z<0.7$ from the SDSS, we show that \oiii\ and 4.6 micron luminosities correlate roughly linearly over 4 orders of magnitude in luminosity, except the low luminosity end. However, there is substantial scatter in this relation. To gain further insight, we carry out a systematic comparison of how the host galaxy properties of AGN change if the nuclear luminosity is parametrized by 4.6 micron luminosity instead of \oiii\ luminosity. We quantify this change using the {\em partition function} of the total integrated 4.6 micron/\oiii\ line luminosity from type 2 AGN as a function of a variety of host galaxy properties including stellar mass, structural properties such as stellar surface mass density and bulge-to-disk ratio, and indicators of stellar population age and ISM dust content.

We find identical distributions of total 4.6 micron and \oiii\ line luminosity for Seyfert galaxies and IR-bright LINERs, in strong support of the standard Unified Model \citep[see also][]{2012ApJ...758....1L}. We also note that if we divide our sample by optical Eddington parameter or 4.6 micron luminosity scaled by black hole mass and if we repeat the comparisons using the 25\% of the emission coming from the IR and optical sources with the {\em highest accretion rates}, host galaxy properties are also identical.

Finally, we note that we searched the entire SDSS spectroscopic catalogue for AGN that could only be identified as such using \wise\ photometry. We found a total of 96 such systems. A detailed analysis revealed that there was nothing special about these objects: in most of them, the S/N in the \hb\ line was simply too low to allow a reliable BPT classification. One might be tempted to conclude, therefore, that no differences exist between the optical and IR ``view'' of low redshift AGN. In a companion paper (Shao et al. 2013, in preparation), we present new results on close pair counts of AGN as a function of both \oiii\ and 4.6 micron luminosity.

\section*{ACKNOWLEDGMENTS}

We thank the anonymous referee for her/his comments which help to improve the quality of this paper. We also thank Dieter Lutz, Reinhard Genzel, Moshe Elitzur and Sebastian H\"onig for helpful discussions and comments. C.~L. acknowledges the support of the 100 Talents Program of Chinese Academy of Sciences (CAS), Shanghai Pujiang Programme (No. 11PJ1411600), and the exchange program between Max-Planck Society and CAS.

Funding for the SDSS and SDSS-II has been provided by the Alfred P. Sloan Foundation, the Participating Institutions, the National Science Foundation, the U.S. Department of Energy, the National Aeronautics and Space Administration, the Japanese Monbukagakusho, the Max Planck Society, and the Higher Education Funding Council for England. The SDSS is managed by the Astrophysical Research Consortium for the Participating Institutions. The Participating Institutions are the American Museum of Natural History, Astrophysical Institute Potsdam, University of Basel, Cambridge University, Case Western Reserve University, University of Chicago, Drexel University, Fermi National Accelerator Laboratory, the Institute for Advanced Study, the Japan Participation Group, Johns Hopkins University, the Joint Institute for Nuclear Astrophysics, the Kavli Institute for Particle Astrophysics and Cosmology, the Korean Scientist Group, The Chinese Academy of Sciences (LAMOST), the Leibniz Institute for Astrophysics, Los Alamos National Laboratory, the Max-Planck-Institute for Astronomy (MPIA), the Max-Planck-Institute for Astrophysics (MPA), New Mexico State University, Ohio State University, University of Pittsburgh, University of Portsmouth, Princeton University, the US Naval Observatory and the University of Washington.

This publication makes use of data products from the Wide-field Infrared Survey Explorer, which is a joint project of the University of California, Los Angeles, and the Jet Propulsion Laboratory/California Institute of Technology, funded by the National Aeronautics and Space Administration.

This publication makes use of data products from the Two Micron All Sky Survey, which is a joint project of the University of Massachusetts and the Infrared Processing and Analysis Center/California Institute of Technology, funded by the National Aeronautics and Space Administration and the National Science Foundation.

\bibliographystyle{mn2e}
\bibliography{agn}

\label{lastpage}

\end{document}